\documentclass[11pt]{article}                                  % do not modify
\usepackage{epsfig}                  % uncomment this line to include figure(s) 
\begin{document}                                               % do not mofify
\begin{flushright}
LA-UR-17-29533
\end{flushright}
\begin{center}                                                 % do not modify

{\bf                                                           % do not modify
%%%%%%%%%%%%%%%%%%%%%%%%%%%%%%%%%%%%%%%%%%%%%%%%%%%%%%%%%%%%%%%%%%%%%%%%
%% TITLE OF THE ABSTRACT 
%% Replace with your title using only upper case letters
%
Fundamental Physics with Electroweak Probes of Nuclei\\ 
%%%%%%%%%%%%%%%%%%%%%%%%%%%%%%%%%%%%%%%%%%%%%%%%%%%%%%%%%%%%%%%%%%%%%%%%
}                                                             % do not modify 
\vspace{4ex}                                                  % do not modify
%%%%%%%%%%%%%%%%%%%%%%%%%%%%%%%%%%%%%%%%%%%%%%%%%%%%%%%%%%%%%%%%%%%%%%%%
%% AUTHORS
%% Replace authors shown with your authors
%% with the Presenting Author underlined
% 
S. Pastore \\
%%%%%%%%%%%%%%%%%%%%%%%%%%%%%%%%%%%%%%%%%%%%%%%%%%%%%%%%%%%%%%%%%%%%%%%%
\vspace{2ex}                                           % do not modify
{\sl                                                   % do not modify
%%%%%%%%%%%%%%%%%%%%%%%%%%%%%%%%%%%%%%%%%%%%%%%%%%%%%%%%%%%%%%%%%%%%%%%%
%% AFFILIATIONS
%
Theoretical Division, Los Alamos National Laboratory, Los Alamos, NM 87545, USA

%%%%%%%%%%%%%%%%%%%%%%%%%%%%%%%%%%%%%%%%%%%%%%%%%%%%%%%%%%%%%%%%%%%%%%%%
}                                                             % do not modify
\vspace{1ex}                                                  % do not modify
{\small \sl                                                   % do not modify
%%%%%%%%%%%%%%%%%%%%%%%%%%%%%%%%%%%%%%%%%%%%%%%%%%%%%%%%%%%%%%%%%%%%%%%%
%% E-Mail
%% Please fill in the email of the presenting author
%
Contact e-mail:  saori@lanl.gov\\

%%%%%%%%%%%%%%%%%%%%%%%%%%%%%%%%%%%%%%%%%%%%%%%%%%%%%%%%%%%%%%%%%%%%%%%
}                                                             % do not modify
\end{center}                                                  % do not modify
\vspace{5mm}                                                  % do not modify
%%%%%%%%%%%%%%%%%%%%%%%%%%%%%%%%%%%%%%%%%%%%%%%%%%%%%%%%%%%%%%%%%%%%%%%%
%% MAIN BODY
%
%
\abstract{The past decade has witnessed tremendous progress in the theoretical and computational tools
that produce our understanding of nuclei. A number of microscopic calculations of nuclear
electroweak structure and reactions have successfully explained the available experimental data, 
yielding a complex picture of the way nuclei interact with electroweak probes. This achievement
is of great interest from the pure nuclear-physics point of view. But it is of much broader
interest too, because the level of accuracy and confidence reached by these calculations 
opens up the concrete possibility of using nuclei to address open questions in other sub-fields 
of physics, such as, understanding the fundamental properties of neutrinos, or the particle nature of dark matter. 

In this talk, I will review recent progress in microscopic calculations of electroweak properties 
of light nuclei, including electromagnetic moments, form factors and transitions in between low-lying 
nuclear states along with preliminary studies for single- and double-beta decay rates. I
will illustrate the key dynamical features required to explain the available experimental data,
and, if time permits, present a novel framework to calculate neutrino-nucleus cross sections for $A>12$ nuclei.}

\section*{}

The nuclear {\it ab initio} approach aims at describing 
the widest range of nuclear phenomena through interactions 
occurring between nucleons inside the nucleus. In this microscopic
picture, nucleons interact with each other via two- and three-body 
interactions, and with external electroweak probes via couplings
to individual nucleons and to nucleon-pairs (a contribution described by
two-nucleon currents). {\it Albeit} limited to
light nuclei ($A\le12$), Quantum Monte Carlo calculations based on the AV18~\cite{AV18}
two-body and IL7~\cite{IL7} three-body interactions successfully explain available 
experimental data in a broad energy range, 
from the keV regime relevant to astrophysics studies to
the GeV regime where short-range correlations become 
predominant~\cite{Carlson15,Carlson98,review2014}.
These studies yield a rather complex picture of the nucleus
with many-body correlations in both the nuclear 
wave functions and electroweak currents playing
an important role in reaching agreement with the data. 
For example, corrections from two-body electromagnetic
currents are as large as 40\% in the calculated magnetic 
moment of $^9$C~\cite{Pastore12}, while electron scattering experiments 
have demonstrated the requirement of two-body currents in quasi-elastic 
scattering from nuclei, where they enhance the transverse response by 
up to $\sim 40\%$~\cite{Carlson98,review2014}.   

The success of the microscopic picture in explaining the data both qualitatively 
and quantitatively is an important achievement from the nuclear physics point of view. 
But it is of much broader interest too, because the level of accuracy and confidence reached by 
these {\it ab inito} Quantum Monte Carlo calculations opens up the concrete possibility
of using nuclei to address prominent and pressing open questions in nuclear physics
and their connection to fundamental physics quests.

Recently, we addressed the ``$g_A$ problem'', that is the systematic
overprediction ($\sim 20 \%$ in $A\le 18$ nuclei) of Gamow-Teller matrix elements in simplified nuclear
calculations~\cite{Chou93,Engel:2016xgb}.
The overprediction of the calculated matrix elements 
is possibly attributable to the fact that for larger nuclear systems, in order 
for the calculations to be computationally feasible, 
one has to  approximate the {\it ab initio} framework, by, 
{\it e.g.}, leaving out correlations and/or truncate the model space. 
Another approximation that can contribute to the 
manifestation of the ``$g_A$ problem'' is in the adopted
model for the nuclear axial current, which typically neglects many-body terms.
In order resolve this long-standing problem, we performed numerically
exact Quantum Monte Carlo calculations of Gamow-Teller matrix elements 
in $A=6$--$10$ nuclei~\cite{Pastore17}, accounting systematically 
for many-body effects in nuclear interactions and coupling to the axial one-, two- 
and three-body currents derived in chiral effective field theory~\cite{Baroni15,Krebs2016}. 

 \begin{figure}[h]
 \label{fig:f1}
 \begin{center}
 \includegraphics[width=4.2in]{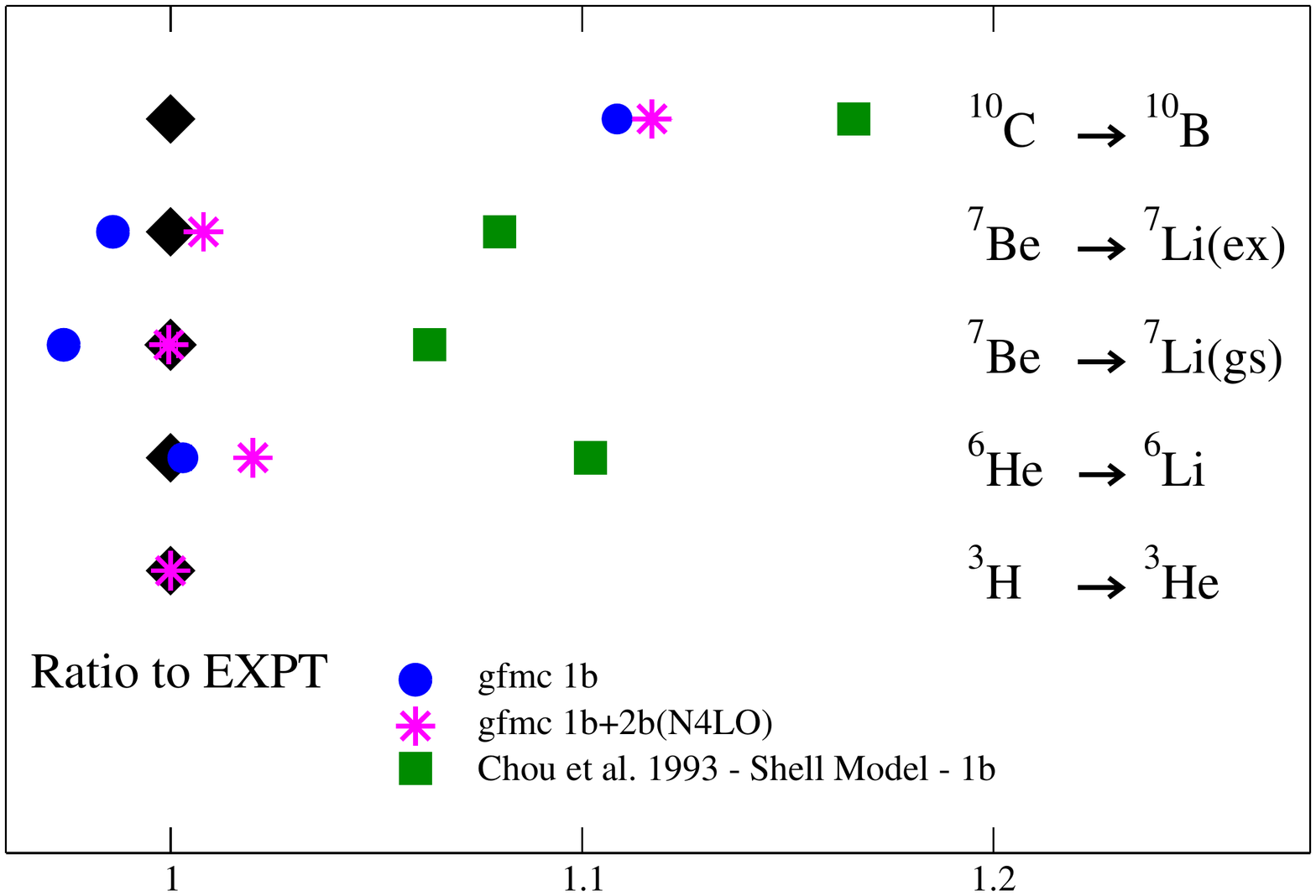}
 \caption{
 (Color online) Ratios of GFMC to experimental values
 of the Gamow-Teller reduced matrix elements  in the $^3$H, $^6$He, $^7$Be, and $^{10}$C weak transitions.  Theory
 predictions (from Ref.~\cite{Pastore17}) correspond to the $\chi$EFT axial current in LO (blue circles) and up to 
 N4LO (magenta stars). Green squares indicate `unquenched' shell model calculations 
 from Ref.~\cite{Chou93} based on the LO axial current. 
 }
  \end{center}
\end{figure}

Our results are summarized in Fig.~1, where we show Green's Function Monte Carlo 
(GFMC)~\cite{Carlson15} calculations of the Gamow-Teller reduced matrix elements 
normalized to the corresponding experimental values in $^3$H, $^6$He, $^7$Be, and 
$^{10}$C weak transitions. The effect of many-body
components in the axial current can be appreciated by comparing blue dots, 
based on the one-body axial current, with the magenta stars which include
in addition to the one-body also two- and three-body terms in the axial current~\cite{Baroni15}.
We find that the effect of many-body currents is negligible, in fact they increase by
only $\sim 2$--$3\%$ the one-body Gamow-Teller contributions.
On the other hand, correlations in the wave functions significantly reduce the 
matrix elements, a fact that can be appreciated by comparing the GFMC (blue
circles in Fig.~1) and the shell model calculations (green squares in
the same figure) from Ref.~\cite{Chou93}, both based on the one-body axial current. 
The reduced matrix elements in the shell model calculations are enhanced by
$\sim 8\%$ ($\sim 17\%$) in $A=6$--$7$ ($A=10$) transitions 
 with respect to the experimental data, while inclusion of
correlations in the nuclear wave functions leads to an agreement with
the data that is excellent in $A=6$--$7$ transitions and at the $10\%$
level in the $^{10}$C weak transition. The discrepancy in this last
case may be attributable to deficiencies in the AV18+IL7 wave functions
of $A=10$ nuclei. In fact, GFMC calculations based on the Norfolk  two- and 
three-nucleon chiral potentials~\cite{NV,Piarulli17} and the 
one-body axial current, bring the  $^{10}$C prediction only $\sim 4\%$ above 
the experimental datum.
These findings suggest that the longstanding ``$g_A$-problem''
may be resolved primarily  by correlation effects. 

While the ``$g_A$-problem'' in single-beta decays, limited to 
the transitions we have studied, seems to be largely attributable 
to missing correlations in the nuclear wave functions, little it is know about how it
propagates at moderate values of momentum transfer.  
This lack of knowledge impacts ongoing world-wide experimental
enterprises aimed at potentially observe neutrinoless double beta 
decay, a process in which two neutrons decay into two protons with the
emission of two electrons but no neutrinos, thus violating the
lepton number conservation by two units. The average momentum transfer in these 
transitions is of the order of 100 MeV~\cite{Engel:2016xgb}, 
a scale that is set by the average distance between the two decaying neutrons. 
Observation of this decay would be a clear signature of physics
beyond the standard model and it will have tremendous consequences on our understanding 
of the Majorana nature of the neutrino and the neutrino mass hierarchy, and potentially 
the observed matter-antimatter asymmetry in the universe. The interpretation of either a 
positive or null experimental result heavily relies on accurate evaluations of nuclear matrix elements. 
The latter are at present characterized by large theoretical uncertainties~\cite{Engel:2016xgb}, 
primarily attributable to the fact that for nuclei relevant to experimental purposes ($A\geq 48$), 
one has to approximate the {\it ab initio} framework by leaving out some correlations.

\begin{figure}[bt]
\centering
 \includegraphics[width=3.5in]{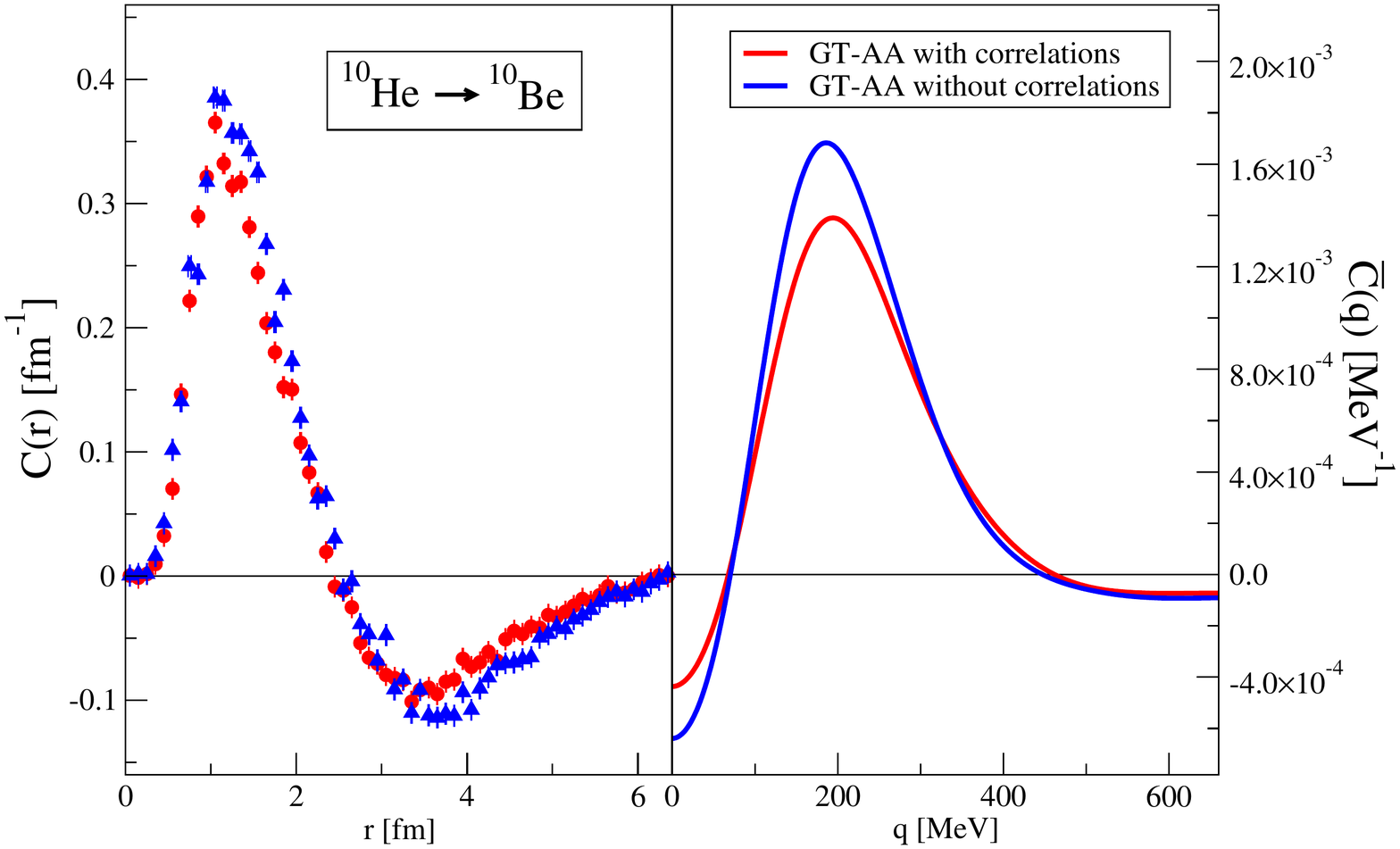} 
 \caption{
 The left (right) panel shows the light Majorana neutrino exchange mechanism (GT-AA)  
 distribution in $r$-space ($q$-space) for the $^{10}$He$\rightarrow^{10}$Be  transition, 
 with and  without ``one-pion-exchange-like'' correlations 
 in the nuclear wave functions. See text and Ref.~~\cite{Pastore:2017ofx}
 for explanation.  
 }
\label{fig:corr10}
\end{figure}

In order to address how the ``$g_A$-problem'' propagates at moderate values of momentum transfer,
we recently performed Quantum Monte Carlo calculations of neutrinoless double beta decay matrix 
elements in $A=6$--$12$  nuclei~\cite{Pastore:2017ofx}. In particular,
we have studied the  effect 
of artificially turning off correlations in 
the nuclear wave functions,  finding a $\sim 10 \%$ increase in the
calculated nuclear matrix elements for the 
light Majorana neutrino exchange mechanism.
This corresponds to having to ``quench'' $g_A$ by $\sim 0.95$
to accommodate for correlation effects. On the other hand,
we saw that in single-beta decay, a zero momentum transfer
observable, the required ``quenching'' of $g_A$ in say, 
the $^{10}$C weak transition evaluated in the shell model, 
is $\sim 0.83$ as one can read off Fig.~1 but comparing the
$A=10$ green square with the experimental datum (black diamond).
These findings may indicate that the $g_A$ ``quenching''
required in calculations based on more approximated nuclear models
(for  $A>12$ nuclei) is larger in single beta decay than in neutrinoless
double beta decays.
In Fig.~2 we show the effect of artificially turning off correlations
in the nuclear wave functions when calculating the neutrinoless double beta decay 
matrix elements induced by the light Majorana neutrino exchange potential. 
In particular, in Fig.~2 
the blue triangles (solid line) in the left (right)
panel represent the $r$-space ($q$-space)  transition
distribution obtained by turning off the correlations
to be compared with the red dots (solid line) 
obtained with the correlated wave function.
The associated matrix element, obtained
by integrating the distribution in $dr$ where
$r$ is the inter-particle distance, undergoes
a $10\%$ increase when correlation are turned off. 

\begin{figure}[bt]
\centering
 \includegraphics[width=4.5in]{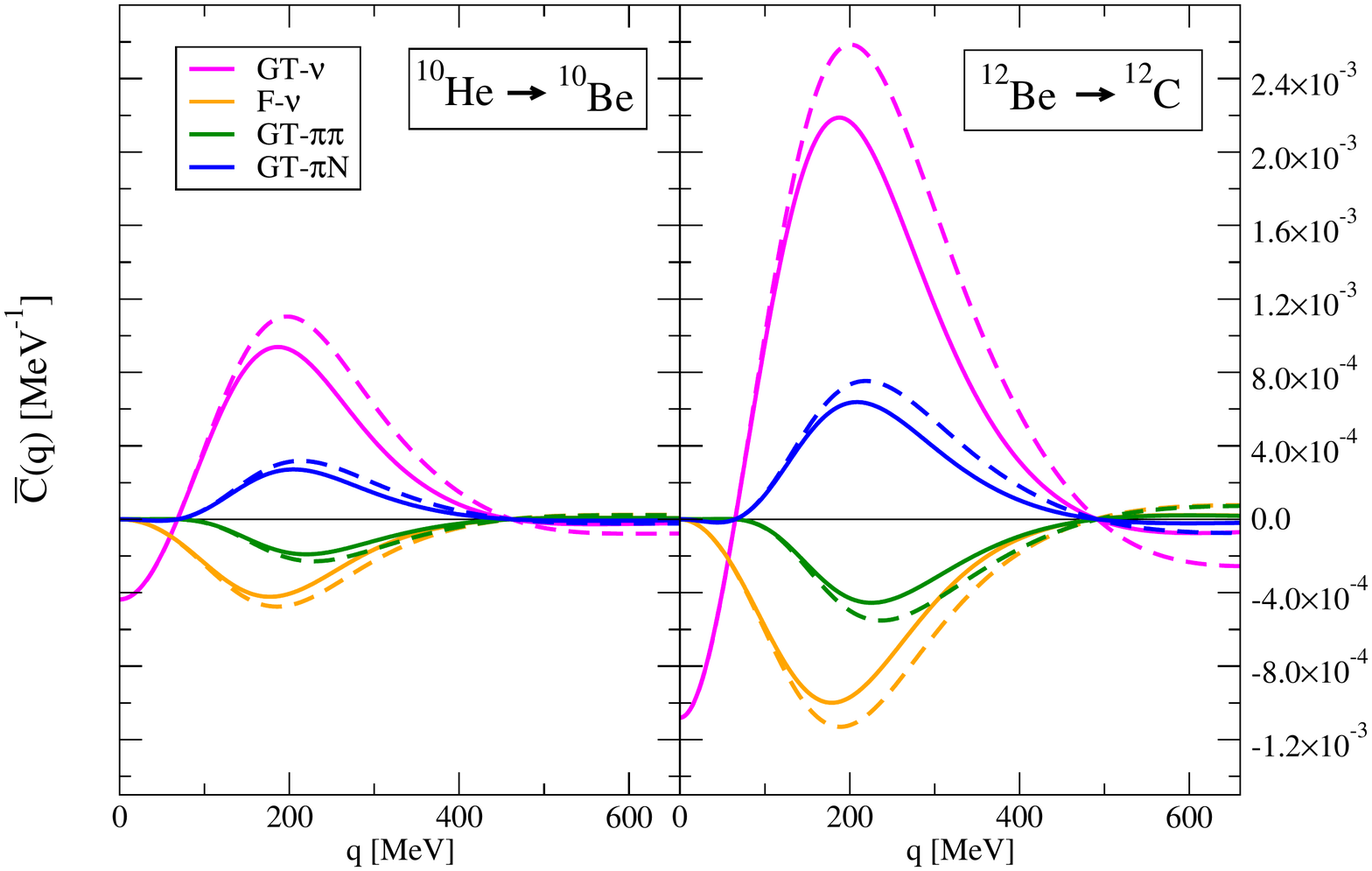} 
 \caption{The GT-$\nu$, F-$\nu$, GT-$\pi\pi$, and GT-$\pi N$ 
 distributions in momentum space for the $^{10}$He$\rightarrow^{10}$Be and $^{12}$Be$\rightarrow^{12}$C
 decays. Solid and dashed lines are obtained, respectively, 
 with and without the inclusion of the momentum dependence in nucleonic form factors.
 See text and Ref.~\cite{Pastore:2017ofx}
 for explanation.}
\label{fig:10vs12q}
\end{figure}

We emphasize that the nuclear systems we studied in Ref.~\cite{Pastore:2017ofx} are
not relevant from the experimental point of view, nevertheless they 
are  interesting and provide us with an extremely useful set of test cases
to benchmark theoretical nuclear models and/or computational  methods.
We studied the matrix elements of light Majorana neutrino exchange potentials~\cite{Bilenky:2014uka,Bilenky:1987ty} 
(denoted by $\nu$ in Fig.~3) as well as  short-distance 
sources (denoted by $\pi\pi$ and $\pi N$ in Fig.~3) of lepton number violation encoded in dimension-seven and -nine 
operators (see Ref.~\cite{Cirigliano:2017djv} and references therein).
These generically lead to Gamow-Teller and Fermi spin-isospin structures,
in Fig.~3 denoted with GT and F, respectively.  
In the left and right panels of Fig.~3 we show the $q$-space transition distributions 
of the  $^{10}$He$\rightarrow^{10}$Be and $^{12}$Be$\rightarrow^{12}$C decays, respectively.
The different height of the peaks is due to the different spatial overlaps 
between an initial diffuse neutron distribution and a final compact proton 
distribution in the case of the $A=10$ transition, and between two compact 
initial neutron and final proton distributions in the $A=12$ transition~~\cite{Pastore:2017ofx}. 
These findings may have implications in our understanding of the 
dynamics entering the neutrinoless double beta decay matrix elements
in nuclei of experimental interest.

\end{document}